Research Article

Eduard Muslimov*, Kjetil Dohlen, Benoit Neichel and Emmanuel Hugot

# Design of pre-optics for laser guide star wavefront sensor for the ELT

**Abstract:** In the present paper, we consider the optical design of a zoom system for the active refocusing in laser guide star wavefront sensors. The system is designed according to the specifications coming from the Extremely Large Telescope (ELT)-HARMONI instrument, the first-light, integral field spectrograph for the European (E)-ELT. The system must provide a refocusing of the laser guide as a function of telescope pointing and large decentring of the incoming beam. The system considers four moving lens groups, each of them being a doublet with one aspherical surface. The advantages and shortcomings of such a solution in terms of the component displacements and complexity of the surfaces are described in detail. It is shown that the system can provide the median value of the residual wavefront error of 13.8–94.3 nm and the maximum value <206 nm, while the exit pupil distortion is 0.26–0.36% for each of the telescope pointing directions.

# 1 Introduction

Anisoplanatism is one of the main limitations in wide-field imaging assisted by single-conjugate adaptive optics (SCAO). It is a consequence of the turbulence volume distribution above the telescope (mainly along the first 20 km

*Corresponding author: Eduard Muslimov, Aix Marseille Univ, CNRS, LAM, Laboratoire d'Astrophysique de Marseille, 38 rue Frédéric Joliot-Curie, Marseille, 13388, France,
e-mail: eduard.muslimov@lam.fr; and Kazan National Research Technical University named after A.N. Tupolev-KAI, 10 K. Marx, Kazan, 420111, Russian Federation
Kjetil Dohlen, Benoit Neichel and Emmanuel Hugot: Aix Marseille Univ, CNRS, LAM, Laboratoire d'Astrophysique de Marseille, 38 rue Frédéric Joliot-Curie, Marseille, 13388, France above the telescope), and it has a significant impact on the images resulting from SCAO correction, severely limiting the size of the corrected field of view. The observation of larger fields of view requires tomographic measurement and correction of the atmospheric turbulence. This technique is based on measuring several guide stars spread across the field of view. The lack of guide stars outside the galactic plane limits the sky coverage of such adaptive optics (AO) systems and calls for the use of artificial guide stars.

Artificial guide stars, referred to as laser guide stars, are created by exciting the sodium layer in the upper atmosphere with a powerful laser. The excited zone in the layer re-emits the light and can be used as a guide star for adaptive optics wavefront sensors. Some limitations appear: (1) With the sodium layer being at a finite-altitude, the laser-assisted AO systems suffer from a cone effect, where the part of the atmosphere crossed by the laser (a cone) is different from the part of the atmosphere crossed by the scientific object (a cylinder). Tomographic reconstruction of the atmospheric volume allows to overcome this limitation. (2) With the sodium layer being at a finite altitude, the distance between the artificial star and the telescope varies with the telescope pointing, leading to a defocus that has to be compensated by means of dynamic refocusing systems.

In this work, we present the design of an optical system that compensates the defocus in a laser guide star wavefront sensor working on an extremely large telescope. This work uses the specifications coming from the HARMONI instrumental requirements. HARMONI [1] is a visible and near-infrared integral field spectrograph, providing the European-extremely large telescope (E-ELT)'s core spectroscopic capability. To get the full sensitivity and spatial resolution gain, HARMONI will work at diffraction-limited scales, thus, requiring the use of AO systems.

The laser guide star wavefront sensor (LGSWFS) is placed between the telescope's pre-focal station (PFS) and the science instrument, housed within a large cryostat. When operating with the lasers, a dichroic beam splitter is inserted into the beam, reflecting the 589-nm sodium wavelength toward the LGSWFS.

The main challenge for the design of the LGSWFS for the E-ELT comes from the large variation in back focal length (BFL), reaching several meters from the viewpoint of the telescope focus. It implies that the wavefront sensor's pre-optics design be re-thought to minimize the translational movements and the size of optical components, while minimizing aberration in a view to maintain the quality of the tomographic wavefront reconstruction. The need to accept variable laser asterism diameters according to Zenith distance is another challenge, although the difficulty of this depends on the design approach chosen.

For HARMONI, several design options for the LGSWFS have been considered. The goal of the current paper is to investigate the possibility of using the zoom lens approach for the LGS projection system of a 40-m class telescope. This technology is highly developed and commonly employed, for instance, in cinematographic zoom optics. Potentially, it can provide the control of fixed image and pupil positions, as well as the pupil size. All the initial assumptions and targeted values are summarized in Table 1 [1–4].

A number of LGS imaging systems with focus compensation were designed and implemented on existing 8–10-m class telescopes [5–7], and a few designs for the 40-m class telescopes are currently under development [8, 9].

**Table 1:** Initial data and boundary conditions for the design.

| LGS | |
|---|---|
| Zenith angle (°) | 0–60 |
| Distance (km) | 80–100 |
| Working wavelength (nm) | 589 |
| Telescope | |
| Focal length (m) | 646.794 |
| Primary mirror diameter (m) | 38 |
| Stop (M4) diameter (m) | 2.254 |
| Stop inclination (°) | 7.75 |
| Detector | |
| Size (pixels) | 800×800 |
| Pixel size (μm) | 24 |
| LGSWFS optics | |
| Design concept | Zoom system consisting of moving axisymmetric lenses, without trombone and/or additional collimators and field lenses |
| Boundary conditions | |
| System length (m) | 1.5–4 |
| System height (m) | <2.5 |
| Distance from the telescope focal plane (m) | ≥1 |
| Lens groups focal length (mm) | >90 |
| Exit pupil diameter (mm) | 19.2 |

For the LGS channel in the GALACSI instrument mounted at the 8-m VLT telescope [5], the back focal distance varies from 93 mm to 175 mm, which corresponds to the LGS distances of 160 km to 85 km, respectively. The similar value for the LGSWFS at the Gemini telescope [6] is 91–182 mm. In both cases, a simple zoom system can be used for the refocusing.

A similar solution was studied for an ELT-class telescope as part of the ATLAS (Advanced Tomography with Laser for AO Systems) project [8]. The proposed design provides a high optical quality across the FoV – from 58.9 nm to 176.7 nm in terms of RMS wavefront error. An interesting feature of this design is the component sizes – the extracting mirror aperture is 440×540 mm, and the first doublet has a diameter of 400 mm.

However, as the development of the ATLAS concept, the laser tomography adaptive optics for the E-ELT has been changed considerably, including different pointing angles of the LGS sources, changes in the pre-optics position with respect to the natural guide star focus. In particular, the HARMONI LGS asterism diameter is much smaller than that of the ATLAS concept, requiring the use of a dichroic beam splitter to separate the laser light from the science light [10, 11]. Therefore, the design solutions made for the ATLAS are not applicable to the HARMONI case.

Other groups also considered the use of zoom lenses, with designs based on different sets of assumptions [12].

## 2 Optical design

The optical design was made in Zemax software, and the optimization and analysis procedures mentioned hereafter rely on the tools proposed by that program. After a preliminary analysis and some trials and errors, it was found that the requirements related to the incoming beam geometry, the optical system dimensions, and its components' main parameters (see Table 1) can be met by the optical system configuration shown in Figure 1. It consists of four moving lens groups and folding mirrors. The lens displacements are required for refocusing and to maintain the aberration correction. The complexity of the lens design is defined by the following factors. On the one hand, it is necessary to account for the beam displacement and also for the high f-number for the first three components. On the other hand, while the optical system operates with monochromatic light, and the FoV is relatively small, it must account for the finite sodium layer thickness. In addition, the optimization should be performed for a number of separate configurations corresponding

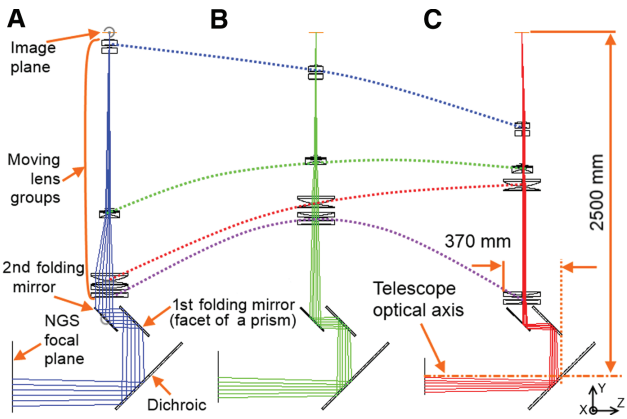

**Figure 1:** General view of the system with components movement. (A) LGS at zenith (90-km altitude), (B) LGS at 45° (127 km altitude), (C) LGS at 60° (180-km altitude). The open gray circles indicate the mirrors' rotation axes (the second mirror is hidden to show the intermediate image plane).

to different distances to the sodium layer and different points along the layer depth.

Each of the components represents an air-spaced doublet made of BK7 glass with an aspherical first surface. As one can note from the diagram, all the lens groups move along a significant distance during the system operation. The total displacements for each of the lens groups are 574.9 mm, 683.5 mm, 388.4 mm and 600.3 mm, respectively. Moreover, the displacement law is obviously nonlinear, and its direction changes with refocusing. These large and complex movements are considered as the main difficulty related to the chosen optical design approach. Nevertheless, the movements are achievable either with four separate motors or with a transmission mechanism like a reduction drive with a few output spindles. Compactness in the $X$ and $Z$ dimensions and flexibility of the design geometry allow to place all the necessary mechanics.

The first folding mirrors from ix LGS channels make up a six-facet prism. It allows to arrange six identical channels side-by-side. The second folding mirror can be rotated to maintain the image centered with a transverse movement of the chief ray. Another rotating mirror should be placed after the image plane to center the exit pupil. The rotation angles of the mirrors are 1.113° and 0.79°, respectively.

## 3 Image and wavefront quality

The optical quality provided by the system is controlled twice, namely, in the image planes and in the exit pupil plane. The spot diagram root-mean square (RMS) radii at the 0° position are 4.0 μm, 15.1 μm and 17.7 μm for the layer's center, its lower and upper edges, respectively. The corresponding values for the 45° position are 0.6, 16.5 and 8.6 μm; for 60°, they are 6.7, 7.1 and 2.0 μm. These numbers are compatible with the Airy disk radius, which is 18.8 μm.

Further, we estimate the wavefront error through the standard Zernike coefficients. The results are shown in Figure 2: the Noll notation is used, and the given layer depth is measured along the plumb line with account for the object tilt change. The tip and tilt are excluded from these plots because they eventually do not affect the wavefront quality. The defocus appears only because of the sodium layer depth and cannot be compensated with a zoom system; so it is also excluded. In the worst case, which takes place for the 0° zenith angle, the RMS wavefront error reaches 206 nm (i.e. 0.35 waves). The corresponding values for 45° and 60° are 154 and 37 nm, respectively. We should note that the influence of the LGSWFS aberrations for the field of view edges is overestimated here. In a real system, they should be weighted according to the sodium intensity distribution model used (see Ref. [13] or [14], for example).

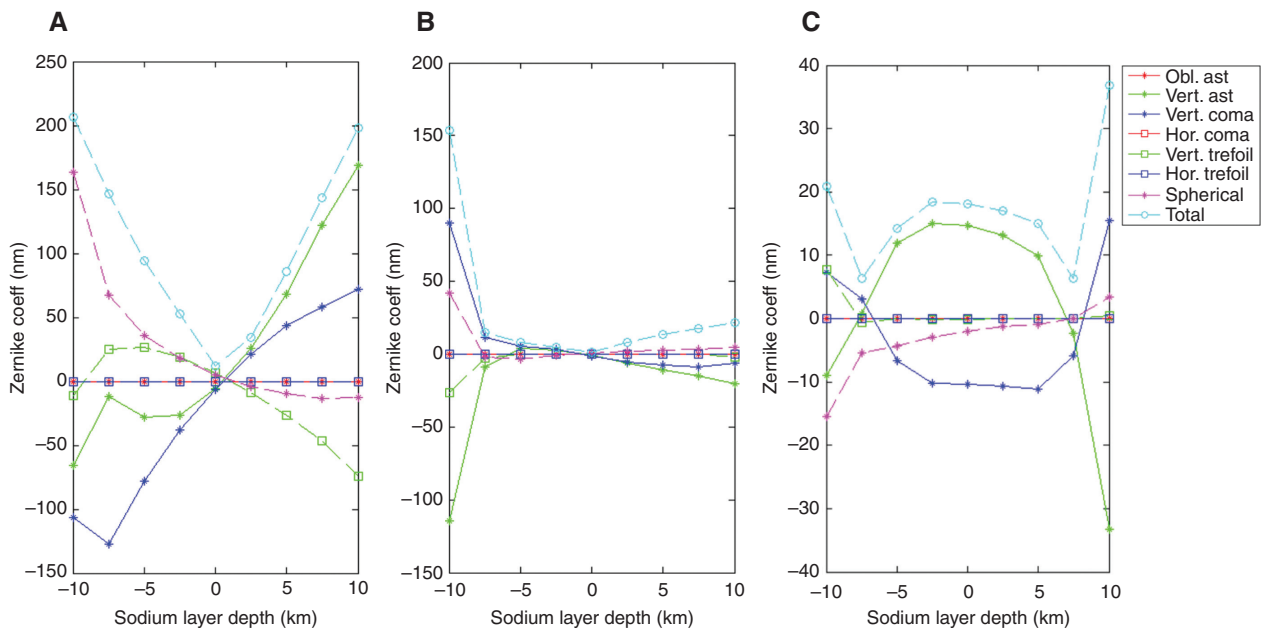

**Figure 2:** Noll Zernike coefficients for the fifth to the 11th orders. (A) The 0° position, (B) The 45° position, (C) The 60° position.

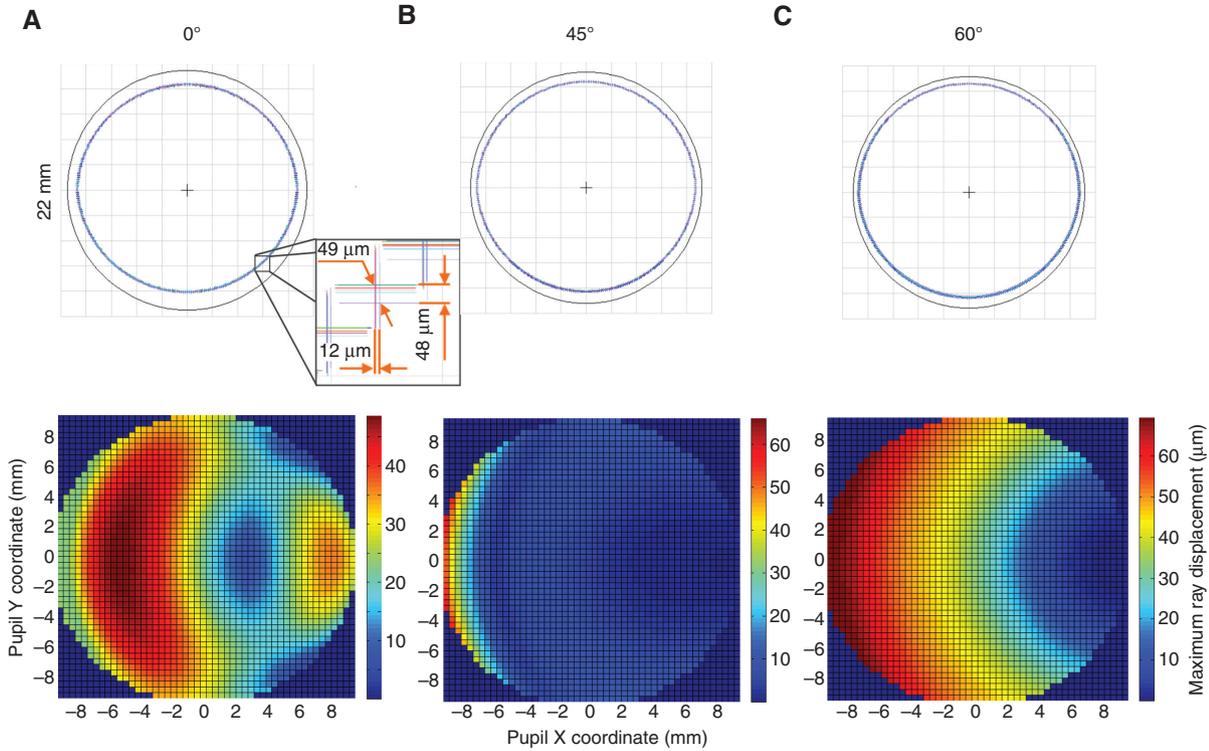

**Figure 3:** Footprint diagrams and ray displacement at the exit pupil plane. (A) LGS at zenith, (B) LGS at 45°, (C) LGS at 60°.

The maintenance of the exit pupil position and size was one of the main boundary conditions during the optimization. The quality and the stability of the pupil image are illustrated in Figure 3. The footprint diagrams in the upper row show that the pupil position is stable. However, the image is characterized by a certain ellipticity, and the ellipse dimensions change for different configurations. The typical ray displacement is shown on the magnified part. Such a displacement (or difference of pupil aberrations) was measured across a grid in each case and plotted in the lower row of Figure 3. The values on these plots correspond to the maximum absolute ray displacement (i.e.

**Table 2:** Exit pupil stability estimation.

| Telescope angle | Maximum absolute ray displacement (µm) | Maximum relative ray displacement (%) | Pupil X size (mm) | Pupil Y size (mm) |
|---|---|---|---|---|
| 0° | 49 | 0.26 | 19.18 | 18.11 |
| 45° | 66 | 0.34 | 19.15 | 18.38 |
| 60° | 70 | 0.36 | 19.24 | 18.61 |

displacement vector length) from a ray corresponding to the center of the FoV. Note that the pupil is located at a constant finite distance.

The quantitative estimation of the pupil size and aberrations are summarized in Table 2. Note that the distortion values given there are relative, so they are calculated for the anamorphic pupil. We should emphasize that changes in pupil diameter between the different telescope positions do not exceed 5.6% and can, therefore, be calibrated.

## 4 Aspherical surfaces analysis

All the aspherical surfaces used in the system are rotationally symmetric. Each asphere was optimized twice, first, in a separate model where a doublet with a calculated

**Table 3:** Aspherical surface main parameters.

| Position | Equation | Max sag | Max asphericity | Clear aperture | RMS slope |
|---|---|---|---|---|---|
| 1st group 1st lens | Even asphere 6th order | 7.271 mm | 125.1 µm | 250.2 mm | 6.3 fr/mm |
| 2nd group 1st lens | Zernike 4th + 11th terms | 1.205 mm | 178.3 µm | 255.2 mm | 9.7 fr/mm |
| 3rd group 1st lens | Q-bfs 0th–4th orders | 9.377 mm | 209.6 µm | 93 mm | 30.9 fr/mm |
| 4th group 1st lens | Zernike 4th + 11th terms | 4.823 mm | 200 µm | 145 mm | 17.3 fr/mm |

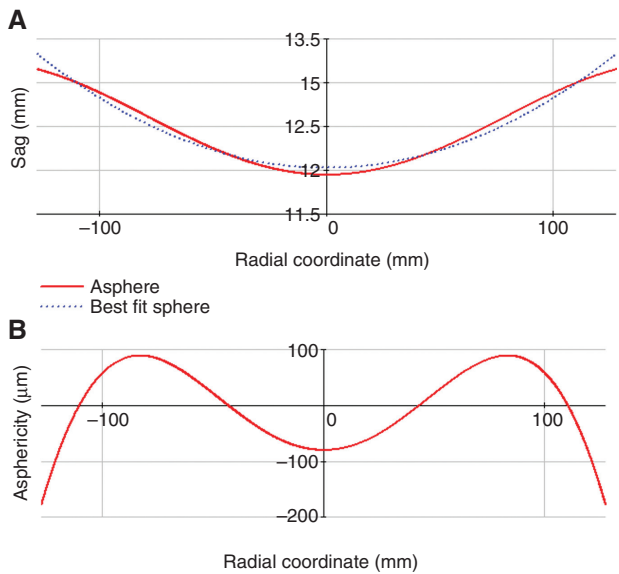

**Figure 4:** Third lens profile. (A) Cross-section, (B) Asphericity. The lens clear aperture is 255.2 mm.

*f*-number was created, second, after insertion of the doublet into the zoom lens system. Different types of equations were tested and compared in terms of optimization convergence speed. It was concluded that the use of the Zernike or Q-bfs [15] polynomials brought some advantages to the optimization process. For instance, the Q-bfs polynomials show the fastest merit decrease speed in the beginning and the smallest value in the end when applied for the design and optimization of the third aspherical doublet.

We should note that there is no commonly adopted method to define the asphere manufacturability, and some of the approaches proposed [16, 17] are valid only for the second-order surfaces or for surfaces without inflection points. In our case, an aspherical surface complexity is estimated through its departure from the best-fit sphere (BFS) and the departure slope to values directly measured with an interferometer, i.e. fringes per millimeter [18, 19]. All the data on the aspheres are summarized in Table 3.

Figure 4 illustrates the profile and asphericity of the third lens, which has the most complex shape.

To conclude, all the surfaces have asphericities, which can be measured with an appropriate precision. It implies that the obtained parameters are within the limits of current technology. The corresponding equations are relatively simple and can be rewritten in a unified form if necessary.

In order to illustrate the reasons leading to these high asphericities as well as to the large component movements, we consider the beam changes after the third aspherical surface (on the fifth lens). The transverse displacement is 19.9 mm for the chief ray, and this value becomes even larger if the finite thickness of the sodium layer is accounted for. Figure 5A shows the beam displacements occurring while the system is refocusing. Figure 5B demonstrates the changes in Zernike coefficients with refocusing (log scale is applied for a better visibility). It is clear that the difference can be as large as a few orders of magnitude. It should be kept in mind that the chart includes unavoidable aberrations of the preceding lenses. The demonstrated beam reshaping features are inherent for LGSWFS at 40-m class telescope, and they explain the design complexities described above.

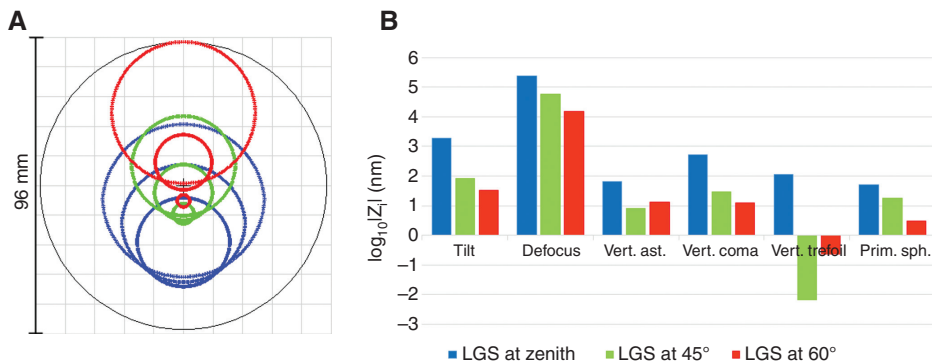

**Figure 5:** Change of the beam at the third aspherical surface. (A) Footprint diagram, (B) Noll Zernike coefficients.

## 5 Conclusions

A zoom lens system designed for use with the LGSWFS for an E-ELT instrument like is demonstrated. This system allows to compensate the LGS image defocusing when the telescope moves from a zenith position to a 60° zenith angle and the associated change in line-of-sight distance to the sodium layer from 90 km to 180 km. The corresponding change in back-focal distance compared with the natural star focal plane is from 4.65 m to 2.32 m. The proposed system consists of four moving lens groups and folding mirrors. Two of the mirrors move as well to perform beam centering.

The presented design solution clearly demonstrates the existing trade-off between complexity and performance in optical system of this type. A few advantages of the system can be noted.

1. Only axisymmetric aspherical surfaces, spheres, and planes are used that would simplify the manufacturing of the optical components.
2. For the aspheres, the maximum displacement from the best-fitting sphere does not exceed 210 μm, and the maximum clear aperture diameter is 256 mm. Both of these values are within capacities of the current technologies.
3. The system shows a relatively high optical performance. Particularly, the intermediate images are diffraction limited, and the RMS wavefront aberrations excluding defocusing and tilt do not exceed 206 nm.
4. The system width is 370 mm for one LGS channel, compatible with the reserved space at the telescope site.

However, the system has a number of significant shortcomings.

1. The system has many moving components – four translating lens doublets and two rotating mirrors. Moreover, the maximum lens displacement is as large as 684 mm, and the movement law is complex and nonlinear. This should be considered as the main disadvantage of this design. We suppose this is inherent, to some extent, for such kind of zoom systems because there are no other possibilities to compensate for the large transverse shift of the incoming beam (accounting for the assumptions made in the beginning).
2. Beams from neighboring LGS are too close to each other, which can cause overlapping on the six-facet prism. It implies that the first air gap should be increased in a more advanced design.
3. There is no real LGS image in front of the system, which would be strongly desirable for performing in-lab testing. The only solution to this problem, like for the previous point, is to increase the length of the first air gap.

To conclude, the developed system consists of relatively simple optical components but is characterized by a large and complex displacement of these components. The current baseline design for the HARMONI instrument is based on an alternative approach, which makes use of more complex optical surfaces to deal with the beam transverse shift and a mirror trombone to compensate for the defocusing. The baseline will be presented in detail elsewhere. If it is found to be technologically difficult, then, the proposed concept could be considered at a more detailed level including tolerance analysis, opto-mechanical design, and other points. However, we believe that the present design study could be useful for developers of auxiliary optics for the ELT-class telescopes.

**Acknowledgments:** Eduard Muslimov acknowledges the support of the European Research Council through the funding of the H2020 – ERC-STG ICARUS – 678777 program. The authors also thank the HARMONI consortium for the statement of this design problem. We also thank Johan Kosmalski for the fruitful discussions and sharing data related to the pre-optics design.

**Eduard Muslimov**
Aix Marseille Univ, CNRS, LAM, Laboratoire d'Astrophysique de Marseille, 38 rue Frédéric Joliot-Curie, Marseille, 13388, France, **eduard.muslimov@lam.fr**; and Kazan National Research Technical University named after A.N. Tupolev-KAI, 10 K. Marx, Kazan, 420111, Russian Federation

Eduard Muslimov is a postdoctoral researcher at LAM. He received his PhD degree in Optical Technology from the Kazan National Research Technical University in 2013. His research interests include complex optical system design, freeform optics, spectral instruments, diffractive optics, and holographic optical elements.

**Kjetil Dohlen**
Aix Marseille Univ, CNRS, LAM, Laboratoire d'Astrophysique de Marseille, 38 rue Frédéric Joliot-Curie, Marseille, 13388, France

Kjetil Dohlen is the head of the Optics Department at LAM. He received his doctoral degree in Applied Optics from the Imperial College London in 1993. As an optical designer and a system engineer, he contributed to such projects as SPHERE instrument at VLT and SPIRE at Herschel telescope and others.

**Benoit Neichel**
Aix Marseille Univ, CNRS, LAM, Laboratoire d'Astrophysique de Marseille, 38 rue Frédéric Joliot-Curie, Marseille, 13388, France

Benoit Neichel is a CR1 researcher at LAM. He received his PhD degree from ONERA in 2008. Then, he worked as an instrument scientist for the Gemini MCAO system (GeMS) and post-doctorant at LAM. His primary research field is adaptive optics.

**Emmanuel Hugot**
Aix Marseille Univ, CNRS, LAM, Laboratoire d'Astrophysique de Marseille, 38 rue Frédéric Joliot-Curie, Marseille, 13388, France

Emmanuel Hugot is a French astrophysicist, head of the R&D group in optics and instrumentation at LAM. He is an ERC grantee, Co-I of two activities in the H2020 OPTICON network, CNRS bronze Medal, and MERAC prize from the European Astronomical Society. He co-authored seven patents, 16 papers, and 50 conference proceedings and, today, leads the European activity on curved detectors together with partners at CEA-LETI.